\documentclass[letterpaper,11pt]{article}

\usepackage{fullpage}
\usepackage{latexsym}
\usepackage{amsmath}
\usepackage{amssymb}
\usepackage{amsfonts}

\def\01{\{0,1\}}

\newcommand{\eps}{\varepsilon}
\newcommand{\EQ}{\mbox{EQ}}

\newcommand{\HAM}{\mbox{HAM}}
\newcommand{\ket}[1]{|#1\rangle}
\newcommand{\bra}[1]{\langle#1|}

\newcommand{\norm}[1]{\mbox{$\parallel{#1}\parallel$}}
\newcommand{\inp}[2]{\langle{#1}|{#2}\rangle} 

\newtheorem{definition}{Definition}
\newtheorem{theorem}{Theorem}
\newtheorem{lemma}{Lemma}

\newtheorem{corollary}{Corollary}

\newenvironment{proof}
{\noindent {\bf Proof. }}
{{\hfill $\Box$}\\
 \smallskip}

\begin{document}

\title{Strengths and Weaknesses of Quantum Fingerprinting}
\author{Dmitry Gavinsky\thanks{University of Calgary.}
 \and Julia Kempe\thanks{CNRS \&\ LRI, Univ.~de Paris-Sud, Orsay. Supported in part by ACI S\'ecurit\'e Informatique
SI/03 511 and ANR AlgoQP grants of the French Research Ministry, and also partially supported by the European
Commission under the Integrated Projects RESQ, IST-2001-37559 and Qubit Applications (QAP) funded by the IST
directorate as Contract Number 015848.}
 \and Ronald de Wolf\thanks{CWI, Amsterdam. Supported by a Veni grant
from the Netherlands Organization for Scientific Research (NWO) and also partially supported by the European Commission
under the Integrated Projects RESQ, IST-2001-37559 and Qubit Applications (QAP) funded by the IST directorate as
Contract Number 015848.}}
\date{}
\maketitle

\begin{abstract}
We study the power of quantum fingerprints in the simultaneous message passing (SMP)
setting of communication complexity. Yao recently showed how to simulate,
with exponential overhead, classical shared-randomness SMP protocols by
means of quantum SMP protocols without shared randomness ($Q^\parallel$-protocols).
Our first result is to extend Yao's simulation to the strongest possible model:
every many-round quantum protocol with unlimited shared entanglement
can be simulated, with exponential overhead, by $Q^\parallel$-protocols.
We apply our technique to obtain an efficient $Q^\parallel$-protocol for a
function which cannot be efficiently solved through more restricted simulations.
Second, we tightly characterize the power of the quantum fingerprinting technique by making a connection to
arrangements of homogeneous halfspaces with maximal margin. These arrangements have been well studied in computational
learning theory, and we use some strong results obtained in this area to exhibit weaknesses of quantum fingerprinting.
In particular, this implies that for almost all functions, quantum fingerprinting protocols
are exponentially \emph{worse} than classical deterministic SMP protocols.
\end{abstract}

\section{Introduction}

\subsection{Setting}

This paper studies the power of quantum fingerprinting protocols in communication complexity. In the simultaneous
message passing (SMP) setting, Alice and Bob hold inputs $x$ and $y$, respectively, and each send a message to a third
party, usually called the ``referee''.  The referee holds no input himself, but is supposed to infer some function
$f(x,y)$ from the messages he receives. The goal is to minimize the amount of communication sent for the worst-case
input $x,y$. In this model there is no direct communication between Alice and Bob themselves, unlike in the standard
model of one-way or multi-round two-party communication complexity. The SMP model is arguably the weakest setting of
communication complexity that is still interesting.

We will consider SMP quantum protocols where Alice sends a $q$-qubit state $\ket{\alpha_x}$, Bob sends a $q$-qubit
state $\ket{\beta_y}$, and the referee does the 2-outcome ``swap test''~\cite{bcww:fp}. This test outputs 0 with
probability
$$
\frac{1}{2}+\frac{|\inp{\alpha_x}{\beta_y}|^2}{2}.
$$
Estimating this probability is tantamount to estimating the absolute value of the inner product
$\inp{\alpha_x}{\beta_y}$. They repeat this $r$ times in parallel, the referee uses the $r$ bits that are the outcomes
of his $r$ swap tests to estimate $|\inp{\alpha_x}{\beta_y}|$, and bases his output on this estimate. We will call such
protocols ``repeated fingerprinting protocols''.

A quantum protocol of this form can only work efficiently if we can
ensure that $|\inp{\alpha_x}{\beta_y}|^2\leq \delta_0$ whenever $f(x,y)=0$ and $|\inp{\alpha_x}{\beta_y}|^2\geq
\delta_1$ whenever $f(x,y)=1$.  Here $\delta_0<\delta_1$ should be reasonably far apart, otherwise $r$ would have to be
too large to distinguish the two cases with high probability.
A statistical argument shows that $r=\Theta(1/(\delta_1-\delta_0)^2)$ is necessary and sufficient for this.
In total, such a protocol uses $2qr=O(q/(\delta_1-\delta_0)^2)$ qubits of communication.
Generally a protocol is considered ``efficient'' if its communication cost is polylogarithmic in the input length. Even
though quantum fingerprinting is a restricted model, it is the only technique we know to get interesting quantum
protocols in the SMP model.

A bit of notation before we get into the study of
quantum fingerprinting: we use $R^{\parallel}(f)$ to denote the minimal
cost among all classical SMP protocols that compute $f$ with error probability
at most $1/3$ on all inputs. Replacing superscript `$\parallel$' by `1',
or removing this superscript altogether, give respectively one-way and
multi-round communication complexity in the standard two-party model without the referee.
Adding superscripts `pub' or `ent' indicates that Alice and Bob share
unlimited amounts of shared randomness or shared entanglement.
These shared resources do not count towards the communication cost.
Replacing `$R$' by `$Q$' gives the variants of these measures where the
communication consists of qubits instead of classical bits.

\subsection{Strengths of quantum fingerprinting}

Quantum fingerprints have surprising power.
They were first used by Buhrman et~al.~\cite{bcww:fp} to show
$Q^\parallel(\EQ)=O(\log n)$ for the $n$-bit equality function.
In contrast, it is known that $R^\parallel(\EQ)=\Theta(\sqrt{n})$
\cite{ambainis:3computer,newman&szegedy:1round,babai&kimmel:simultaneous},
while $R^{\parallel,pub}(\EQ)=O(1)$.
Subsequently, Yao~\cite{yao:qfp} showed that
$$
Q^\parallel(f)=2^{O(R^{\parallel,pub}(f))}\log n.
$$
In particular, if $R^{\parallel,pub}(f)=O(1)$ then $Q^\parallel(f)=O(\log n)$.
The quantum fingerprinting protocol for equality is a special case of this result.
Yao's exponential simulation can be extended to relational problems,
and recently Gavinsky et al.~\cite{gkrw:identification} showed that it is
essentially optimal by exhibiting a relational problem $P_1$ for which
$R^{\parallel,pub}(P_1)=O(\log n)$ and $Q^\parallel(P_1)=\Omega(n^{1/3})$.
Whether there exist exponential gaps for functional problems remains open.

In this paper we show that Yao's simulation can be extended far beyond
classical SMP protocols. Given any bounded-error two-party quantum protocol
with $q$ qubits of communication, no matter how many rounds of communication,
and no matter how much entanglement it starts with, we show how to construct
a repeated quantum fingerprinting protocol that communicates $2^{O(q)}\log n$
qubits and computes the same function with small error probability.
In symbols:
$$
Q^\parallel(f)=2^{O(Q^{ent}(f))}\log n.
$$
Thus, the exponential simulation still works even if we add interaction,
quantum communication, and entanglement to the $R^{\parallel,pub}$-model
that Yao considered.  When we restrict to simulating $R^{\parallel,pub}$-protocols,
we get a bound that is quadratically better than Yao's.
A similar quadratic improvement over Yao's has been obtained independently
by Golinsky and Sen~\cite{golinsky&sen:qfp}.

Actually, the vectors that we construct for our quantum simulation can also be used to obtain a classical SMP protocol
with shared randomness and $O(r)$ bits of communication ($r$ being the number of repetitions
of the quantum protocol), as follows. Alice and Bob use their shared randomness to
pick an $O(1)$-dimensional random subspace and each projects her/his vector onto that space and renormalizes.
The expectation of the inner product of the two projected vectors equals their original inner
product.
They send the resulting $O(1)$-dimensional vectors to the referee in sufficient precision
($O(\log r)$ bits per entry suffices), and repeat this $O(r)$ times to
approximate the inner product between the original vectors with sufficient precision.
Hence our construction implies Shi's result~\cite{shi:tensornorms}
$$
R^{\parallel,pub}(f)=2^{O(Q^{ent}(f))}.
$$
This is not too surprising, because our derivation of the appropriate vectors
(fingerprints) from the $Q^{ent}$-protocol is inspired by some of the techniques
in Shi's paper---though we avoid his use of tensor norms.

The fact that our simulation has exponential overhead
is unfortunate but unavoidable. For instance, for Raz's function~\cite{raz:qcc}
we have $Q(f)=O(\log n)$ via a two-round protocol while it is easy to see
that any quantum fingerprinting protocol needs to communicate $n^{\Omega(1)}$ qubits:
by the argument of the last paragraph, a quantum fingerprinting protocol implies a classical
shared-randomness protocol of roughly the same complexity, and Raz proved that all
classical protocols for his problem require $n^{\Omega(1)}$ bits of communication.
Despite the exponential overhead, our simulation still gives nontrivial efficient
$Q^\parallel$-protocols when simulating protocols with $O(\log\log n)$
quantum communication and much shared randomness or entanglement.
We give an example in Section~\ref{ssecsimexample}.

\subsection{Characterization and weaknesses of quantum fingerprinting}

The results above show some of the strengths of quantum fingerprinting protocols.
What about its weaknesses?  For instance, is it possible that quantum SMP protocols
based on repeated fingerprinting are equal in power to arbitrary quantum SMP protocols?
In Section~\ref{secfpcharacterization} we show that for most functions they are much weaker.

Our main tool is a tight characterization of quantum fingerprinting systems in terms of the optimal margin achievable
by realizations of the computational problem via an arrangement of homogeneous halfspaces (Theorem \ref{th:final}). The
latter mouthful has been well studied in machine learning, and forms the basis of maximal-margin classifiers and
support vector machines. This connection between quantum fingerprints and these embeddings is straightforward,
but allows us to tap into some
of the strong theorems known about such margins, particularly a result of Forster~\cite{forster:probcc} and its recent
strengthening by Linial et~al.~\cite{Linial:sign}. The upshot is that repeated quantum fingerprinting protocols are
exponentially worse than general quantum and even classical SMP protocols for almost all functions.

This three-way connection between quantum communication complexity, margin complexity, and learning theory
allows us to make other connections as well.
For example, good learning protocols give good lower bounds on margins, which give new
upper bounds for repeated fingerprinting protocols. In the other direction,
an efficient multi-round quantum protocol for some Boolean function implies lower bounds
on the margin of the corresponding matrix. We give an example of this in Section~\ref{sseclowermargin}.
Finally, since our positive result above relates quantum fingerprinting to general $Q^{ent}$-complexity,
we can also use known results about margin complexity to obtain some new lower bounds on $Q^{ent}(f)$.
We explore the latter direction in Section~\ref{sseclowerqent}.
There we show $Q^{ent}(f)=\Omega(\log(1/\gamma(f))$, where $\gamma(f)$ is the ``maximal margin''
among all embeddings of $f$.  This bound was independently obtained by Linial and Shraibman~\cite{Linial:disc}
in a recent manuscript, which also shows the beautiful new result that margin complexity
and \emph{discrepancy} are linearly related.

\section{Simulating Arbitrary Quantum Protocols}

In this section we show how to extend Yao's simulation from
classical SMP protocols with shared randomness to multi-round quantum
protocols with shared randomness (Section~\ref{ssecrandomness}),
and then even to arbitrary multi-round quantum protocols with shared \emph{entanglement}
(Section~\ref{ssecentanglement}).

\subsection{Simulating shared-randomness multi-round quantum protocols}\label{ssecrandomness}

Let $f:\01^n\times\01^n\rightarrow\01$ be a communication complexity problem.
Our construction also works for promise functions, but for simplicity we describe it here for a total function.
Let $P$ be the $2^n\times 2^n$ matrix of acceptance probabilities of
a bounded-error quantum protocol for $f$.  We first assume the protocol communicates
$q$ qubits and doesn't use prior shared entanglement or shared randomness.
It is known~\cite{yao:qcircuit,kremer:thesis} that we can decompose
$P=AB^\dagger$ where $A,B$ are $2^n\times 2^{2q-2}$ matrices, each of whose entries has absolute value at most 1,
and $B^\dagger$ is the conjugate transpose of $B$.
Let $a(x)$ be the $x$-th row of $A$ and $b(y)$ be the $y$-th row of $B$.
Then for all $x,y$ we have
$$
P(x,y)=\inp{a(x)}{b(y)}\mbox{ and }\norm{a(x)},\norm{b(y)}\leq 2^{q-1}.
$$
Now consider a quantum protocol that uses shared randomness.
By Newman's theorem~\cite{newman:random}, we can assume without loss of generality
that the shared random string $r$ is picked uniformly from a set $R$ of $O(n)$ elements.
Then we can decompose
$$
P=\frac{1}{|R|}\sum_{r\in R} P_r,
$$
where $P_r$ is the matrix of probabilities if we run the protocol with shared string $r$.
Each $P_r$ induces vectors $a_r(x), b_r(y)$ as above, and we have
$$
f(x,y)\approx P(x,y)=\frac{1}{|R|}\sum_{r\in R}\inp{a_r(x)}{b_r(y)},
$$
where `$\approx$' means that $f(x,y)$ and $P(x,y)$ differ by at most the error probability of the protocol.
Define pure $(q+\log n+O(1))$-qubit states as follows
$$
\ket{\alpha_x}=\frac{1}{\sqrt{|R|}}\sum_{r\in R}\ket{r}\otimes\frac{\ket{a_r(x)}+\sqrt{2^{2q-2}-\norm{a_r(x)}^2}\ket{{\rm junk}_a}}{2^{q-1}}
$$
and
$$
\ket{\beta_y}=\frac{1}{\sqrt{|R|}}\sum_{r\in R}\ket{r}\otimes\frac{\ket{b_r(y)}+\sqrt{2^{2q-2}-\norm{b_r(y)}^2}\ket{{\rm junk}_b}}{2^{q-1}}
$$
where `${\rm junk}_a$' and `${\rm junk}_b$' are distinct special basis states.
Note that
$$
\inp{\alpha_x}{\beta_y}=\frac{1}{|R|}\sum_{r\in R}\frac{\inp{a_r(x)}{b_r(y)}}{2^{2q-2}}=\frac{1}{2^{2q-2}}P(x,y).
$$
Using these states gives a repeated quantum fingerprinting protocol that computes $f$ with small error
and sends $O(2^{8q}\log n)$ qubits of communication, without shared randomness.

\begin{theorem}
$Q^{\parallel}(f)=O(2^{8 Q^{pub}(f)}\log n)$.
\end{theorem}

Note that we put $\log n$ instead $q+\log n$ for the last factor.
That is clearly correct if $q<(\log n)/8$; and if $q\geq(\log n)/8$
then the righthand side is more than $n$, which is a trivially true upper bound on $Q^{\parallel}(f)$.

We can get a better exponent in the case of classical one-way protocols.
Suppose Alice's classical message is $c=R^{1,pub}(f)$ bits.
Let $a_r(x)\in\01^{2^c}$ have a 1 only in the coordinate corresponding to
the message Alice sends given input $x$ and random string $r$.
Let $b_r(y)\in\01^{2^c}$ be 1 on the messages of Alice that lead Bob to output 1 (given $y$ and $r$).
Then $P_r(x,y)=\inp{a_r(x)}{b_r(y)}$, $\norm{a_r(x)}=1$ and $\norm{b_r(y)}\leq\sqrt{2^c}$.
The above fingerprinting construction now gives a protocol with $O(2^{2c}\log n)$ qubits.

\begin{theorem}
$Q^{\parallel}(f)=O(2^{2 R^{1,pub}(f)}\log n)$.
\end{theorem}

Analogously we can simulate classical shared-randomness SMP protocols.
Suppose Alice's messages are $c\leq R^{\parallel,pub}(f)/2$ bits long.
This gives rise to a repeated fingerprinting protocol with
$O(2^{2 c}\log n)$ qubits of communication: define $a_r(x)$ as before and let
$b_r(y)\in\01^{2^c}$ be 1 on the possible messages $a$ of Alice that would lead the referee
to accept given $a$ and the message Bob would send (on his input $y$ and random string $r$).
This bound is quadratically better than Yao's simulation of classical SMP protocols.

\begin{theorem}
$Q^{\parallel}(f)=O(2^{R^{\parallel,pub}(f)}\log n)$.
\end{theorem}

\subsection{Simulating shared-entanglement multi-round quantum protocols}\label{ssecentanglement}

Now consider the case where our multi-round quantum protocol uses $q$ qubits
of communication and some entangled starting state.
Our proof for this most general case is inspired by Shi's result
$R^{\parallel,pub}(f)=2^{O(Q^{ent}(f))}$~\cite[Theorem~1.2]{shi:tensornorms}.
The following lemma is due to Razborov~\cite[Proposition~3.3]{razborov:qdisj}
and is similar to earlier statements in~\cite{yao:qcircuit,kremer:thesis}.
It can be proved by induction on $q$.

\begin{lemma}[Kremer-Razborov-Yao]
Let $\ket{\Psi}$ denote the (possibly entangled) starting state of the protocol.
For all inputs $x$ and $y$, there exist linear operators $A_h(x),B_h(y)$, $h\in\01^{q-1}$,
each with operator norm $\leq 1$, such that the acceptance probability of the protocol is
$$
P(x,y)=\norm{\displaystyle\sum_{h\in\01^{q-1}} (A_h(x)\otimes B_h(y))\ket{\Psi}}^2.
$$
\end{lemma}

We will derive vectors $a(x)$ and $b(y)$ from this characterization.
Assume without loss of generality that the prior entanglement is
$$
\ket{\Psi}=\sum_{e\in E}\lambda_e\ket{e}\ket{e},
$$
with $\{\ket{e}\}$ an orthonormal set of states and $\sum_e\lambda_e^2=1$.
Note that $|E|$ may be huge. Now we can write
$$
P(x,y) = \norm{\displaystyle\sum_{h\in\01^{q-1}} (A_h(x)\otimes B_h(y))\ket{\Psi}}^2
       = \sum_{h,h',e,e'}\lambda_{e'}\bra{e}A_h(x)^\dagger A_{h'}(x)\ket{e'}\cdot\lambda_e\bra{e}B_h(y)^\dagger B_{h'}(y)\ket{e'}.
$$
Define $a(x)$ to be the $|E|^2 2^{2q-2}$-dimensional vector with complex entries
$\lambda_{e'}\bra{e}A_h(x)^\dagger A_{h'}(x)\ket{e'}$, indexed by tuples $(h,h',e,e')$,
and similarly define $b(x)$ with entries $\lambda_e\bra{e}B_h(y)^\dagger B_{h'}(y)\ket{e'}$.
Then
$$
P(x,y)=\inp{a(x)}{b(y)}.
$$
Using that the set of $\ket{e}$-states is an orthonormal set in the space in
which $A_h(x)^\dagger A_{h'}(x)\ket{e'}$ lives, and the fact that
$\norm{A_h(x)^\dagger A_{h'}(x)}\leq\norm{A_h(x)}\cdot\norm{A_{h'}(x)}\leq 1$ we have
$$
\norm{a(x)}^2  =  \sum_{h,h',e,e'}\lambda_{e'}^2|\bra{e}A_h(x)^\dagger A_{h'}(x)\ket{e'}|^2
               \leq  \sum_{h,h',e'}\lambda_{e'}^2\norm{A_h(x)^\dagger A_{h'}(x)\ket{e'}}^2
               \leq  \sum_{h,h',e'}\lambda_{e'}^2 = 2^{2q-2}.
$$
Similarly $\norm{b(y)}\leq 2^{q-1}$.

The norms and inner products of the $a(x)$ and $b(y)$ vectors are thus as before.
It remains to reduce their dimension $D=|E|^2 2^{2q-2}$, which may be very large.
For this we use the Johnson-Lindenstrauss lemma (proved
in~\cite{johnson&lindenstrauss}, see e.g.~\cite{dasgupta&gupta:jl} for a simple proof).

\begin{lemma}[Johnson \&\ Lindenstrauss]\label{jllemma}
Let $\eps>0$ and $d\geq 4\ln(N)/(\eps^2/2-\eps^3/3)$.
For every set $V$ of $N$ points in $\mathbb{R}^D$ there exists a map
$p:\mathbb{R}^D\rightarrow\mathbb{R}^d$ such that for all $u,v\in V$
$$
(1-\eps)\norm{u-v}^2 \leq \norm{p(u)-p(v)}^2 \leq (1+\eps)\norm{u-v}^2.
$$
\end{lemma}

To get the above map $p$, it actually suffices to project the vectors onto a random
$d$-dimensional subspace and rescale by a factor of $\sqrt{D/d}$.
With high probability, this approximately preserves all distances.
Note that if the set $V$ includes the 0-vector, then also the norms
of all $v\in V$ will be approximately preserved. Since
$$
\inp{u}{v}=\frac{\norm{u}^2+\norm{v}^2-\norm{u-v}^2}{2},
$$
the map $f$ also approximately preserves the inner products between
all pairs of vectors in $V$, if $\eps$ is sufficiently small.

We assume for simplicity that our vectors $a(x)$ and $b(y)$ are real.
Let our set $V$ contain all $a(x)$ and $b(y)$ as well as the 0-vector (so $N=2\cdot 2^n+1$).
Applying the Johnson-Lindenstrauss lemma with $\eps=1/(10\cdot 2^{2q})$ and
$d=O(\log(N)/\eps^2)=O(n 2^{4q})$ gives us $d$-dimensional vectors
$p(a(x))$ and $p(b(y))$ of norm at most $2^q$ such that
$$
\left| \inp{p(a(x))}{p(b(y))} - \inp{a(x)}{b(y)} \right| \leq 1/10.
$$
We fix these vectors once and for all before the protocol starts;
note that we are not using shared randomness in the protocol itself.\footnote{Using
shared randomness gives us the result
$R^{\parallel,pub}(f)=2^{O(Q^{ent}(f))}$ of~\cite[Theorem~1.2]{shi:tensornorms}.}

Now define quantum states in $d+2$ dimensions by
$$
\ket{\alpha_x}=\frac{\ket{p(a(x))}+\sqrt{2^{2q}-\norm{p(a(x))}^2}\ket{{\rm junk}_a}}{2^q}
$$
and
$$
\ket{\beta_y}=\frac{\ket{p(b(y))}+\sqrt{2^{2q}-\norm{p(b(y))}^2}\ket{{\rm junk}_b}}{2^q}.
$$
Note that
$$
\inp{\alpha_x}{\beta_y}=\frac{\inp{p(a(x))}{p(b(y))}}{2^{2q}}\approx\frac{\inp{a(x)}{b(y)}}{2^{2q}}=\frac{1}{2^{2q}}P(x,y).
$$
Hence these states form a repeated fingerprinting protocol with fingerprints
of $\log(d+2)= O(q+\log n)$ qubits and $O(2^{8q})$ repetitions.

\begin{theorem}\label{thm:4}
$Q^{\parallel}(f)=O(2^{8 Q^{ent}(f)}\log n)$.
\end{theorem}

\subsection{An example problem}\label{ssecsimexample}

Here we apply Theorem \ref{thm:4} to obtain an efficient SMP protocol for a particular problem;
we do not know how to obtain an efficient protocol for this problem without using Theorem~\ref{thm:4}. More precisely, we give an
example of a Boolean function $f$ for which there exists a 4-round quantum protocol that uses $q=O(\log\log n)$ qubits
of communication and $O(\log n)$ bits of shared randomness. Our simulation implies the existence of an efficient
quantum SMP protocol for $f$:
$$
Q^{\parallel}(f)\leq 2^{O(\log\log n)}\log n=(\log n)^{O(1)}.
$$
The problem uses many small copies of Raz's 2-round communication problem from~\cite{raz:qcc},
and is defined as follows.
\begin{quote}
{\bf Alice's input:} string $x\in\01^k$, unit vectors
$v_1,\ldots,v_k\in\mathbb{R}^m$, and $m/2$-dimensional subspaces $S_1,\ldots,S_k$ of $\mathbb{R}^m$\\
{\bf Bob's input:} string $y\in\01^k$, and $m$-dimensional unitaries $U_1,\ldots,U_k$\\
{\bf Promise:} $|x\oplus y|=k/\log\log k$, and either\\
\hspace{1em} ($f=0$) $U_iv_i\in S_i$ for each $i$ where $x_i\oplus y_i=1$, or\\
\hspace{1em} ($f=1$) $U_iv_i\in S_i^\perp$ for each $i$ where $x_i\oplus y_i=1$
\end{quote}
As stated this is a problem with continuous input, but we can easily approximate
the entries of the vectors, unitaries, and subspaces by $O(\log m)$-bit numbers.
Thus the input length is $n=O(k m^2\log m)$ and we choose $m=\log k$.

Here's a simple 4-round protocol for this problem.
First, Alice and Bob use shared randomness to pick $O(\log\log k)$ indices
$i\in[k]$. Alice sends the corresponding $x_i$ to Bob, Bob sends
the corresponding $y_i$ to Alice. They pick the first index $i$ such that
$x_i\oplus y_i=1$ (there will be such an $i$ in their $O(\log\log k)$-set
with high probability). Then Alice sends $v_i$ to Bob as a $\log m$-qubit state.
Bob applies $U_i$ and sends back the result $U_iv_i$, which is another $\log m$ qubits.
Alice measures with subspace $S_i$ versus $S_i^\perp$ and outputs the result (0 or 1).
The overall communication is $2\log\log k + 2\log m=O(\log\log n)$.

Note that we need both shared randomness and multi-round quantum communication
to achieve $Q^{pub}(f)=O(\log\log n)$,
and hence to achieve $Q^\parallel(f)=(\log n)^{O(1)}$ via our simulation.
In contrast, Yao's simulation from~\cite{yao:qfp} cannot give us an
efficient $Q^\parallel$-protocol. This is because every classical many-round protocol
(including SMP shared-randomness ones) for even one instance of Raz's
problem needs about $\sqrt{m}\approx\sqrt{\log n}$ bits of communication~\cite{raz:qcc}.
The same lower bound then also holds for the classical SMP model with shared randomness.
Hence the best $Q^\parallel$-protocol that Yao's simulation could give is
$2^{O(\sqrt{m})}\log n\approx 2^{\sqrt{\log n}}$.

\section{Characterization of Quantum Fingerprinting}\label{secfpcharacterization}

As mentioned, all nontrivial and nonclassical quantum SMP protocols known
are based on repeated fingerprinting. Here we will analyze the power of
protocols that employ this technique, and show that it is closely related
to a well studied notion from computational learning theory.
This addresses the 4th open problem Yao states in~\cite{yao:qfp}.
In particular, we will show that such quantum fingerprinting protocols
cannot efficiently compute many Boolean functions for which there is
an efficient classical SMP protocol.

\subsection{Embeddings and realizations}

We now define two geometrical concepts.

\begin{definition}
Let $f:{\cal D}\rightarrow\01$, with ${\cal D}\subseteq X\times Y$,
be a (possibly partial) Boolean function.
Consider an assignment of unit vectors $\alpha_x\in\mathbb{R}^d$,
$\beta_y\in\mathbb{R}^d$ to all $x\in X$ and $y\in Y$.

This assignment is called a \emph{$(d,\delta_0,\delta_1)$-threshold embedding of $f$}
if $|\inp{\alpha_x}{\beta_y}|^2\leq\delta_0$ for all $(x,y)\in f^{-1}(0)$
and $|\inp{\alpha_x}{\beta_y}|^2\geq\delta_1$ for all $(x,y)\in f^{-1}(1)$.

The assignment is called a
\emph{$d$-dimensional realization of $f$ with margin $\gamma>0$} if
$\inp{\alpha_x}{\beta_y}\geq\gamma$ for all $(x,y)\in f^{-1}(0)$ and
$\inp{\alpha_x}{\beta_y}\leq-\gamma$ for all $(x,y)\in f^{-1}(1)$.
\end{definition}

Our notion of a ``threshold embedding'' is essentially
Yao's~\cite[Section~6, question~4]{yao:qfp}, except that we square
the inner product instead of taking its absolute value,
since it is the square that appears in the swap test's probability.
Clearly, threshold embeddings and repeated fingerprinting protocols
are essentially the same thing (with fingerprints of $\log d$ qubits,
and $O(1/(\delta_1-\delta_0)^2)$ repetitions).
The notion of a ``realization'' is computational learning theory's
notion of the realization of a concept class by an arrangement
of homogeneous halfspaces.

These two notions are essentially equivalent:

\begin{lemma}\label{lemembedding}
If there is a $(d,\delta_0,\delta_1)$-threshold embedding of $f$,
then there is a $(d^2+1)$-dimensional realization of $f$ with margin
$\gamma=(\delta_1-\delta_0)/(2+\delta_1+\delta_0)$.

Conversely, if there is a $d$-dimensional realization of $f$ with margin $\gamma$,
then there is a $(d+1,\delta_0,\delta_1)$-threshold embedding of $f$
with $\delta_0=(1-\gamma)^2/4$ and $\delta_1=(1+\gamma)^2/4$.
\end{lemma}

\begin{proof}
Let $\alpha_x,\beta_y$ be the vectors in a $(d,\delta_0,\delta_1)$-threshold
embedding of $f$.  For $a=(\delta_1+\delta_0)/(2+\delta_1+\delta_0)$,
define new vectors $\alpha'_x=(\sqrt{a},\sqrt{1-a}\cdot\alpha_x\otimes\alpha_x)$
and $\beta'_y=(\sqrt{a}, -\sqrt{1-a}\cdot\beta_y\otimes\beta_y)$.
These are unit vectors of dimension $d^2+1$.
Now
$$
\inp{\alpha'_x}{\beta'_y}=a-(1-a)|\inp{\alpha_x}{\beta_y}|^2.
$$
If $(x,y)\in f^{-1}(1)$, then $|\inp{\alpha_x}{\beta_y}|^2\geq\delta_1$
and hence
$\inp{\alpha'_x}{\beta'_y}\leq a-(1-a)\delta_1=-\gamma.$
Similarly, $\inp{\alpha'_x}{\beta'_y}\geq\gamma$ for $(x,y)\in f^{-1}(0)$.

For the converse, let $\alpha_x,\beta_y$ be the vectors in a
$d$-dimensional realization of $f$ with margin $\gamma$.
Define new $(d+1)$-dimensional unit vectors
$\alpha'_x=(1,\alpha_x)/\sqrt{2}$ and $\beta'_y=(1,-\beta_y)/\sqrt{2}$. Now
$$
|\inp{\alpha'_x}{\beta'_y}|^2=\frac{1}{4}\left(1-\inp{\alpha_x}{\beta_y}\right)^2.
$$
If $(x,y)\in f^{-1}(1)$, then $\inp{\alpha_x}{\beta_y}\leq-\gamma$
and hence
$|\inp{\alpha'_x}{\beta'_y}|^2\geq\frac{1}{4}\left(1+\gamma\right)^2=\delta_1.$
A similar argument shows
$|\inp{\alpha'_x}{\beta'_y}|^2\leq\frac{1}{4}\left(1-\gamma\right)^2=\delta_0$
for $(x,y)\in f^{-1}(0)$.
\end{proof}

The tradeoffs between dimension $d$ and margin $\gamma$ have been well
studied~\cite{forster:probcc,fklmss:relations,fsss:optmargins,Linial:sign}.
In particular, we can invoke a very strong
bound on the best achievable margin of realizations due to very recent work by
Linial et~al.~\cite[Section~3.2]{Linial:sign} (our $\gamma$ is their $1/mc(M)$).

\begin{theorem}[Linial et al.]\label{th:linial}
For $f:X\times Y\rightarrow\01$, define the $|X|\times|Y|$-matrix $M$ by $M_{xy}=(-1)^{f(x,y)}$. Every realization of
$f$ (irrespective of its dimension) has margin $\gamma$ at most $$\gamma\leq \frac{K_G \cdot \norm{M}_{\ell_\infty
\rightarrow \ell_1}}{|X|\cdot|Y|},$$ where the norm $\norm{M}_{\ell_\infty \rightarrow \ell_1}$ is given by
$\norm{M}_{\ell_\infty \rightarrow \ell_1}=\sup_{\|v\|_{\ell_\infty}=1} \norm{Mv}_{\ell_1}$ and $1<K_G<1.8$ is
Grothendieck's constant.
\end{theorem}

This bound is the strongest known upper bound for the margin of a sign matrix. It strengthens the previously known
bound due to Forster~\cite{forster:probcc}:

\begin{corollary}[Forster]
Every realization of $f$ (irrespective of its dimension) has margin $\gamma$ at most $\gamma\leq
\norm{M}/\sqrt{|X|\cdot|Y|}$, where $\norm{M}$ is the operator norm (largest singular value) of $M$.
In particular, if $f:\01^n\times\01^n\rightarrow\01$ is the inner product function, then
$\norm{M}=\sqrt{2^n}$ and hence $\gamma\leq 1/\sqrt{2^n}$.
\end{corollary}

Combining this with Lemma~\ref{lemembedding}, we see that
a $(d,\delta_1,\delta_0)$-threshold embedding of the inner product
function has $\delta_1-\delta_0=O(1/\sqrt{2^n})$.
In repeated fingerprinting protocols, we then need $r\approx 2^n$
different swap tests to enable the referee to reliably distinguish
0-inputs from 1-inputs!
Hence if we consider the function $f(x,y)$ defined by the inner product
function on the first $\log n$ bits of $x$ and $y$, there is an efficient
classical SMP protocol for $f$ (Alice and Bob each send their first
$\log n$ bits), but even the best quantum fingerprinting protocol needs
to send $\Omega(n)$ qubits. The same actually holds for almost
all functions defined on the first $\log n$ bits.  This indicates
an essential weakness of quantum fingerprinting protocols.

In general, the preceding arguments show that we cannot have an efficient repeated fingerprinting protocol if $f$
cannot be realized with large margin. If the largest achievable margin is $\gamma$, the protocol will need
$\Omega(1/\gamma^2)$ copies of $\ket{\alpha_x}$ and $\ket{\beta_y}$.
We now show that this lower bound is close to optimal. Consider a
realization of $f: X \times Y \rightarrow \{0,1\}$ with maximal margin $\gamma$. Its vectors may have very high
dimension, but nearly the same margin can be achieved in fairly low dimension if we use the Johnson-Lindenstrauss lemma
\cite{johnson&lindenstrauss}.  Assume without loss of generality that $|X| \geq |Y|$ and let $n=\log |X|$.

\begin{lemma}
A $D$-dimensional realization of $f$ with margin $\gamma$ can be converted into an $O(n/\gamma^2)$-dimensional
realization of $f$ with margin $\gamma/2$.
\end{lemma}

Using Lemma~\ref{lemembedding}, this gives us a $(d,\delta_1,\delta_0)$-threshold embedding of $f$ with
$d=O(n/\gamma^2)$, $\delta_0=(1-\gamma/2)^2/4$ and $\delta_1=(1+\gamma/2)^2/4$. Note that $\delta_1-\delta_0=\gamma/2$.
This translates directly into a repeated fingerprinting protocol with states $\ket{\alpha_x}$ and $\ket{\beta_y}$ of
$d$ dimensions, hence $O(\log(n/\gamma^2))$ qubits, and $r=O(1/\gamma^2)$. For example, if $f$ is equality then
$\gamma$ is constant, which implies an $O(\log n)$-qubit repeated fingerprinting protocol for equality (of course, we
already had one with $r=1$). In sum:

\begin{theorem}\label{th:final}
For $f:X\times Y\rightarrow\01$ with $2^n=|X| \geq |Y|$, define the $|X|\times|Y|$-matrix $M$ by
$M_{xy}=(-1)^{f(x,y)}$, and let $\gamma$ denote the largest margin among all realizations of $M$. There exists a
repeated fingerprinting protocol for $f$ that uses $r=O(1/\gamma^2)$ copies of $O(\log(n/\gamma^2))$-qubit states.
Conversely, every repeated fingerprinting protocol for $f$ needs $\Omega(1/\gamma^2)$ copies of its $\ket{\alpha_x}$
and $\ket{\beta_y}$ states.
\end{theorem}

\subsection{Application: getting margin lower bounds from communication protocols}\label{sseclowermargin}

The connection between repeated fingerprinting and maximum margin of a realization can be exploited in the reverse
direction as well, by deriving new lower bounds on margin complexity from known communication protocols.
Yao~\cite{yao:qfp} considered the following Hamming distance problem on $n$-bit strings $x$ and $y$:
\begin{quote}
$\HAM^{(d)}_n(x,y)=1$ iff the Hamming distance between $x$ and $y$ is $\Delta(x,y)\leq d$.
\end{quote}
For $d=0$, this is just the equality problem.
Yao showed $R^{\parallel,pub}(\HAM^{(d)}_n)=O(d^2)$ (actually, a better
classical protocol may be derived from the earlier paper~\cite{fimnsw:multiparty}).
We can derive a threshold embedding directly from Yao's classical construction
in~\cite[Section~4]{yao:qfp}.  There, the length of the messages sent by
the parties is $m=\Theta(d^2)$.
The referee accepts only if the Hamming distance between the messages
is below a certain threshold $t=\Theta(m)$.
Let $a_{rx}$ be Alice's message on random string $r$ and input $x$,
$a_{rxi}$ be the $i$-th bit of this message, and similarly for Bob.
Again we may assume $r$ ranges over a set of size $n'=O(n)$~\cite{newman:random}.
Yao shows that for uniformly random $r$ and $i$,
$\Pr[a_{rxi}=b_{ryi}]\leq t/m-\Theta(1/d)$ if $\Delta(x,y)\leq d$,
and $\Pr[a_{rxi}=b_{ryi}]\geq t/m+\Theta(1/d)$ if $\Delta(x,y)>d$.
Here $t/m=\Theta(1)$.
Now define the following $(\log(n')+2\log(d)+1)$-qubit states:
$$
\ket{\alpha_x}=\frac1{\sqrt{mn'}}
 \sum_r\ket{r}\sum_{1\le i\le m}\ket{i}\ket{a_{rxi}}
\mbox{ \ and \ \ }
\ket{\beta_y}=\frac1{\sqrt{mn'}}
 \sum_{r}\ket{r}\sum_{1\le i\le m}\ket{i}\ket{b_{ryi}}.
$$
Then
$$
\inp{\alpha_x}{\beta_y}=\frac1{mn'}
 \sum_{r}\sum_{1\le i\le m}\delta_{a_{rxi},b_{ryi}}=
 \Pr[a_{rxi}=b_{ryi}].
$$
This is a threshold embedding of $\HAM^{(d)}_n$ with $\delta_1-\delta_0=\Theta(1/d)$,
so the margin complexity of this problem is $\gamma(\HAM^{(d)}_n)=\Omega(1/d)$.
We have not found this result anywhere else in the literature on maximum margin realizations
and believe it is novel.

\subsection{Application: a margin-based lower bound on $Q^{ent}(f)$}\label{sseclowerqent}

Let us consider again the unit vectors (a.k.a.~quantum states)
$\alpha_x$ and $\beta_y$ constructed in Section~\ref{ssecentanglement}
from a quantum protocol for function $f$ with $q=Q^{ent}(f)$ qubits of communication.
These states form a $(d,\delta_0,\delta_1)$-threshold embedding of $f$
with $\delta_1-\delta_0=\Theta(2^{-4q})$.
By Lemma~\ref{lemembedding}, this in turn implies that the maximal achievable margin
among all realizations of $f$ is $\gamma(f)=\Omega(2^{-4q})$, which translates into
a lower bound on quantum communication complexity in terms of margins:

\begin{theorem}
$Q^{ent}(f)\geq \frac{1}{4}\log(1/\gamma(f))-O(1)$.
\end{theorem}

Since almost all $f$ have exponentially small maximal margin~\cite[Section~5]{Linial:sign},
it follows that almost all $f$ have linear communication complexity even for multi-round protocols with
unlimited prior entanglement. As far as we know, this is a new result (albeit not a very surprising one).

The last theorem has been independently obtained by Linial and Shraibman~\cite{Linial:disc}. Even more interestingly,
they actually showed a linear relation between margin complexity $1/\gamma(f)$ and \emph{discrepancy}. Hence they
extend the discrepancy lower bound to $Q^{ent}(f)$. It was already known to hold for $Q(f)$ without
entanglement~\cite{kremer:thesis}.

\section{Discussion}

Our simulation is relevant for the longstanding open question regarding the power of quantum entanglement in
communication complexity: how much can we reduce communication complexity by giving the parties access to unlimited
amounts of EPR-pairs? No good upper bounds are known on the largest amount of entanglement (shared EPR-pairs) that is
``still useful''. This is in contrast to the situation with shared randomness,
where Newman's theorem shows that in the standard one-round or multi-round setting, $O(\log n)$
shared coin flips suffice~\cite{newman:random}, and hence shared randomness can save at most $O(\log n)$ communication.%
\footnote{In fact, Jain et~al.~\cite{jrs:messagecompression} show that Newman's blackbox-type proof,
which keeps the protocol the same and just reduces the set of random strings to $O(n)$ elements,
cannot be lifted to the quantum setting to get a significant reduction in the amount of entanglement used.}
Like Shi's result~\cite{shi:tensornorms}, our result does not give an upper bound on the amount of prior
entanglement that is needed, but it does imply that adding large amounts of prior
entanglement can reduce the communication no more than exponentially.

An interesting direction is to tap into the vast literature on maximal-margin classification and support vector
machines (SVM's) to find more natural communication problems having efficient quantum fingerprinting protocols.
Currently, the only natural and nontrivial example we have of this is the equality problem from~\cite{bcww:fp} and its
variations in Section~\ref{secfpcharacterization}. Every learning problem involving a concept class $\cal C$ over the set
of $n$-bit strings corresponds to a $|{\cal C}|\times 2^n$ communication complexity problem. If the learning problem
can be embedded with large margin ($\gamma\geq 1/(\log n)^{O(1)}$, say), the communication problem has an efficient
quantum fingerprinting protocol.

A fascinating line of research which combines our main results is the following. Our Theorem \ref{thm:4} together with
the characterization of repeated fingerprinting in Theorem \ref{th:final} opens the possibility to derive new lower
bounds on the maximum margin of a sign matrix. It is sufficient to give an efficient multi-round quantum communication
protocol (even with unlimited pre-shared entanglement) for a Boolean function to show that the corresponding concept
class can be learned efficiently - yet another interesting possibility of proving classical results the quantum way.
Conversely, strong upper bounds on maximum margin, like the one of Linial et al.~in Theorem \ref{th:linial}, give
lower bounds on the communication complexity in the multi-round quantum communication model with unlimited shared
entanglement.

\subsection*{Acknowledgments}
We thank Oded Regev for discussions and very helpful proofreading, and Adi Shraibman and Nati Linial for discussions
regarding their recent work~\cite{Linial:sign} and~\cite{Linial:disc}.

\bibliographystyle{alpha}

\newcommand{\etalchar}[1]{$^{#1}$}

\end{document}